\documentclass[doublecol]{epl2} 

\newcommand{\be}{\begin{equation}}
\newcommand{\ee}{\end{equation}}
\newcommand{\bs}{\begin{subequations}}
\newcommand{\es}{\end{subequations}}

\usepackage{color}

\title{Effects of van der Waals forces and salt ions on the growth of water films on ice and the detachment of CO$_2$ bubbles}
\author{P. Thiyam\inst{1}\and E. R. A. Lima \inst{2} \and O. I. Malyi \inst{3} \and D. F. Parsons \inst{4}  \and S. Y. Buhmann \inst{5,6} \and C. Persson \inst{1,3,7} \and M. Bostr{\"o}m \inst{3,a} }

\institute{  \inst{1} Department of Materials Science and Engineering,
Royal Institute of Technology, SE-100 44 Stockholm, Sweden\\
\inst{2} Programa de P{\'{o}}s-gradua{\c{c}}{\~{a}}o em
Engenharia Qu{\'{i}}mica, Universidade do Estado do Rio de Janeiro,
CEP 20550-013, Rio de Janeiro RJ, Brazil\\
\inst{3}Centre for Materials Science and Nanotechnology,
University of Oslo, P. O. Box 1048 Blindern, NO-0316 Oslo, Norway\\
\inst{4}School of Engineering and IT, Murdoch University,
90 South St, Murdoch, WA 6150, Australia\\
\inst{5}Physikalisches Institut, Albert-Ludwigs-Universit\"at
Freiburg, Hermann-Herder-Str. 3, 79104 Freiburg, Germany\\
\inst{6}Freiburg Institute for Advanced Studies,
Albert-Ludwigs-Universit{\"{a}}t Freiburg, Albertstra{\ss}e 19, 79104
Freiburg, Germany\\
\inst{7}Department of Physics, University of Oslo,
P. O. Box 1048 Blindern, NO-0316 Oslo, Norway\\
\inst{a}Email corresponding authors: thiyam@kth.se; lima.eduardo@gmail.com;  Mathias.Bostrom@smn.uio.no.}
 \shortauthor{P. Thiyam \etal}


\abstract{
{We study the effect of salts on the thickness of wetting films on
melting ice and interactions acting on CO$_2$ bubble near ice-water
and vapor-water interfaces}. Governing mechanisms are the Lifshitz and
the double-layer interactions in {the respective} three-layer
geometries. We demonstrate
that the latter depend on the Casimir--Polder interaction of the salt
ions dissolved in water with the respective ice, vapour and CO$_2$
interfaces, as calculated using different models for their effective
polarizability in water. Significant variation in the predicted thickness of
the equilibrium water film is observed for different salt ions and when
using different models for the ions' polarizabilities. {We find that
CO$_2$ bubbles are attracted towards  ice-water interface
and repelled from the vapor-water interface}. }

\begin{document}

\maketitle

%\section{Introduction}
 
Brine is known to play a significant role in the dissolution and
migration of CO$_2$~\cite{Orr} during the process of carbon
sequestration in deep geological formations~\cite{Benson,Pires}. A number of investigations
were performed to understand the role of viscous, capillary, and
gravity forces in CO$_2$ storage, and to understand the mechanisms of
CO$_2$ movement~\cite{Rochelle,Taku}. We are interested in the {interaction of CO$_2$ bubbles in water near ice-water and vapor-water interfaces}. It is stabilized by repulsive
Lifshitz forces~\cite{Elbaum}. 
Molecular simulations have also
shown that quasi-liquid layers can be formed on ice
surfaces~\cite{Jungwirth}. We predict a marked increase in the
equilibrium water film thickness on melting ice in the presence
of salt ions compared to the case without salt. This could be highly
relevant for ice melting (e.g in permafrost and on Mars where transient salt water has
been reported~\cite{Torres}).

The interesting aspect of the Lifshitz force is that it can be both attractive and repulsive. This has been understood for a long time\,\cite{Dzya,Rich1,Rich}.  It was found in the work of Anderson and Sabiski on films of liquid helium on {calcium fluoride}, and similar molecularly smooth surfaces \,\cite{AndSab}. The films ranged from 10-200 \AA \,\cite{AndSab}. The thickness of the films could be measured to within a few percent in most cases. For the saturated-film measurements the repulsive van der Waals potential was equal to the negative of the gravitational potential \,\cite{AndSab}. A good agreement was found\,\cite{Rich1} between these experimental data and the results from Lifshitz theory. In another more subtle experiment Hauxwell and Ottewill\,\cite{Haux} measured the thickness of films of oil on water near the alkane saturated vapour pressure. For this system, n-alkanes up to octane spread on water. Higher alkanes do not spread. It is an asymmetric system (oil-water-air), and the surfaces are molecularly smooth. The phenomenon depends on a balance of van der Waals forces against the vapor pressure\,\cite{Rich, Haux,NinParBPJ1970}. The net force, as a function of film thickness depends on the dielectric properties of the oils. As demonstrated in ref.\,\cite{Rich} it involves an intricate balance of repulsive and attractive components from different frequency regimes. When the ultraviolet and visible components are exponentially damped by retardation the opposing (repulsive) infrared components take over\,\cite{NinParBPJ1970}.

\begin{figure}
\centerline{\includegraphics[width=7cm]{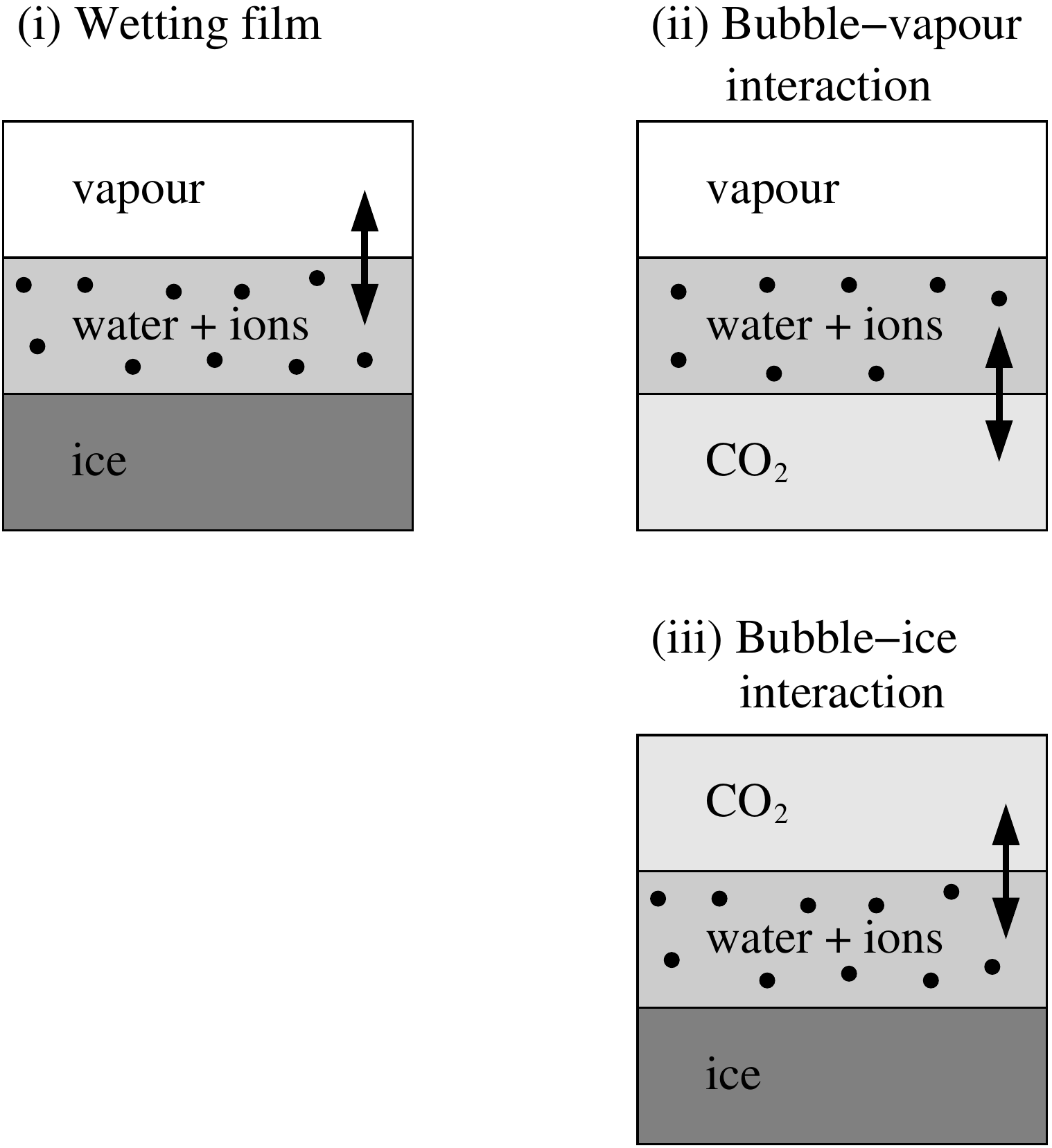}}
\caption{{System under study: (i) The ice--water--vapour system
serves to study the film thickness. (ii) The vapour--water--CO$_2$
system approximates a bubble close to the water--vapour interface
while (iii) the ice--water--CO$_2$ can be similarly used to study the
interaction of the bubble with the water--ice interface.}}
\label{figu0}
\end{figure}

Our {systems} of interest {are} depicted in fig.~\ref{figu0}.  Our aim is to study the combined effects of
the Lifshitz dispersion force and the ion-specific double-layer force
on the film thickness and the{ bubble-interface interactions}. To that end, we
consider three different relevant planar three-layer subsystems: (i)
an ice--water--vapour system can be used to study the film thickness
while (ii) vapour--water--CO$_2$ and (iii) ice--water--CO$_2$ systems
serve to investigate the interaction of the CO$_2$ bubble in water
with the vapour and ice interfaces, respectively. {Here, we assume
that the bubble is very close to the interface, so that
its interaction with the interface can be understood from a planar
geometry via the Derjaguin approximation~\cite{Derjaguin}.  }

In the absence of salt, the retarded Lifshitz {force} turns from
short-range repulsion to long-range attraction in the
ice--water--vapor subsystem with a transition around 2.1 nm
or 3.6 nm
depending on how the dielectric function of ice is modeled.
We use two models ice$_{12}$ and ice$_{13}$ for the dielectric
function of ice based on exerimental data  in the literature, where the
subscript indices indicate the reference numbers listed in ref.~\cite{Elbaum}. In the short range, the double layer interaction adds repulsion to the
ice--water--vapor and CO$_2$--water--vapor {systems} but gives an
attractive contribution to the ice--water--CO$_2$ {system}. This
effect from non-linear van der Waals ion-surface forces is in line
with the work by Dzyaloshinskii et al.~\cite{Dzya,Ninhb}, that predicts changes from
attraction to repulsion depending on the relative values of the
dielectric functions of the three layers comprising the system.

In order to study the effect of salt ions on CO$_2$ bubbles in water and
the equilibrium water film thickness at the melting ice surface, we
apply different models for the effective polarizability of ions in
water. The non-zero effective polarizabilities of ions render
ionic dispersion potentials and hence ion-specific concentration profiles
between the two uncharged interfaces. This gives rise to a non-linear
double-layer contribution~\cite{Ninhb,Yam,Kjellander,BNW} that alters the total force
between the two interfaces.
Note that for two uncharged interfaces there is no double layer interaction in the absence of
finite ion polarizabilities. We explore
the Lifshitz and
double-layer interactions between large
{CO$_2$} bubbles in water with ice--water or vapor--water
interfaces. We discuss how the presence of ions may affect surface
sticking of bubbles as well as the equilibrium thickness of
Lifshitz-repulsion induced water sheets on ice~\cite{Elbaum}.

\begin{figure}
\includegraphics[width=8cm]{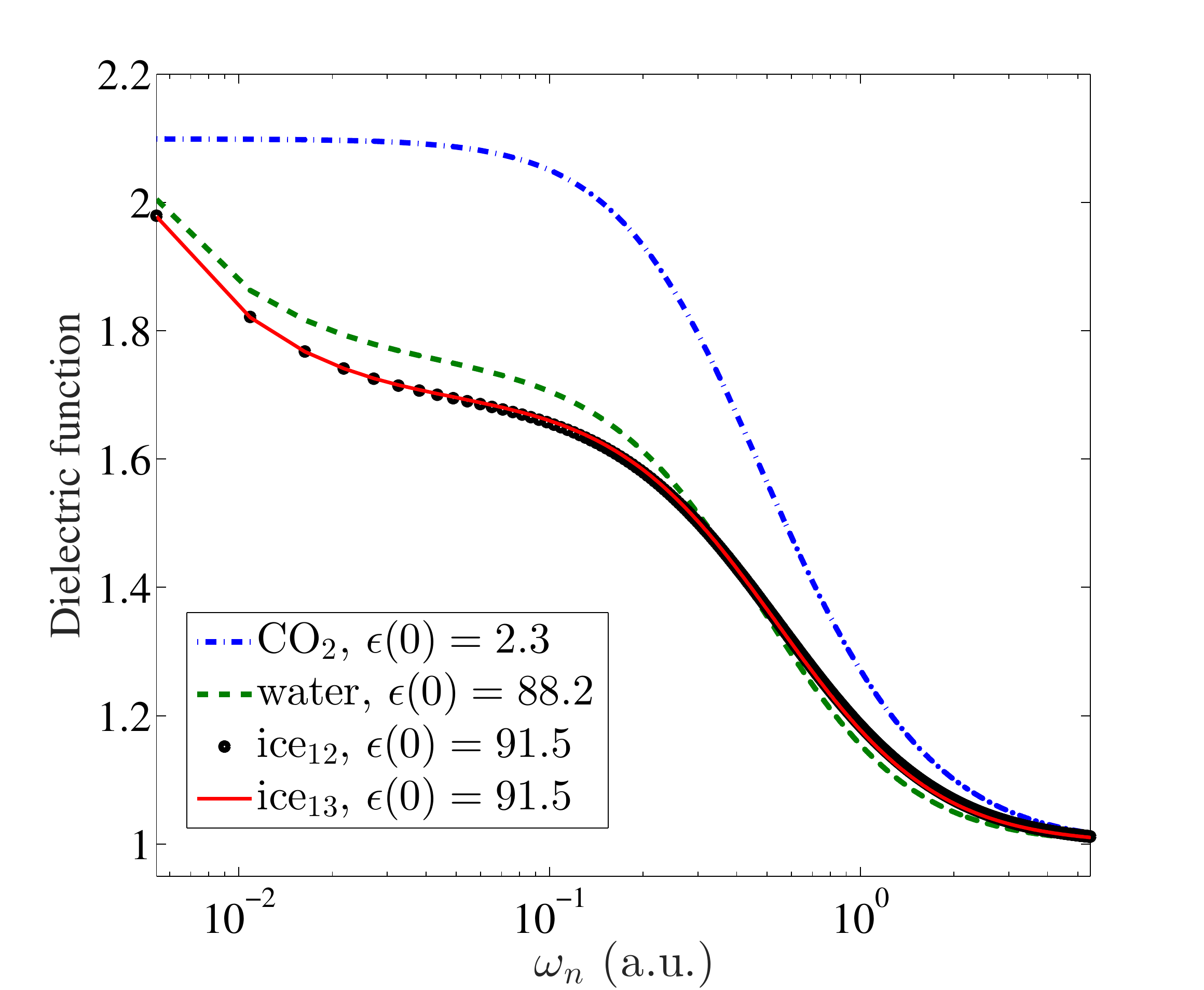}
\caption{(Color online) The dielectric functions of {CO$_2$}, water, ice$_{12}$ and ice$_{13}$  in terms of imaginary Matsubara frequencies. The dielectric constants at zero frequency are also shown in the figure legends.}
\label{figu0b}
\end{figure}

In this study, we consider crystalline CO$_2$ to compute the
dielectric function of the CO$_2$ using the
Perdew-Burke-Ernzerhof functional in density functional theory (DFT). The scissors operator is applied to represent the hybrid functional~\cite{HSE06} band gap of
8.4 eV. We determine the dielectric functions on the imaginary
frequency axis using one version of the Kramers-Kronig dispersion
relation~\cite{Sernelius}. {Such an approach was shown to provide resonable description of water dielectric function\,\cite{French}.} {The
Lifshitz and double layer interactions are not very sensitive to the
dielectric function of CO$_2$. They are much more sensitive to the
dielectric functions of ice and water since these are extremely
similar when the water is in equilibrium with the ice. Because of this
and since for different phases of a material having similar volumes
and electronic properties the dielectric functions are
similar\,\cite{Elbaum, Sasha}, it can be speculated that the results
predicted in this paper remain qualitatively valid for different
CO$_2$ phases (solid, liquid, supercritical).}
Ice$_{12}$ model is based on energy loss spectroscopy~\cite{Daniels}
while ice$_{13}$ model is based on reflectivity spectrum~\cite{Seki}.
As can be observed from fig.~\ref{figu0b}, the two models give very
similar dielectric functions. {These dielectric functions are
for water and ice at the triple point which assumes zero degrees
Celsius and low pressure. However, these dielectric functions have in
the past also been used as approximate dielectric functions for water
in different phases, at other temperatures and pressures, in a
study of van der Waals interactions involving methane gas
hydrates (i.e. a mixture of methane and ice/water
molecules)\,\cite{Bonnefoy}}.
{For different ice models}, only the inclusion of retardation slightly shifts the point  where the Lifshitz {force} becomes zero. The double layer interaction, on the oher hand, is not so sensitive. The two different interactions are not very sensitive to the dielectric function of { CO$_2$} which supports the approximate use of the corresponding crystalline dielectric function.

\begin{figure}
\includegraphics[width=8cm, height=6cm]{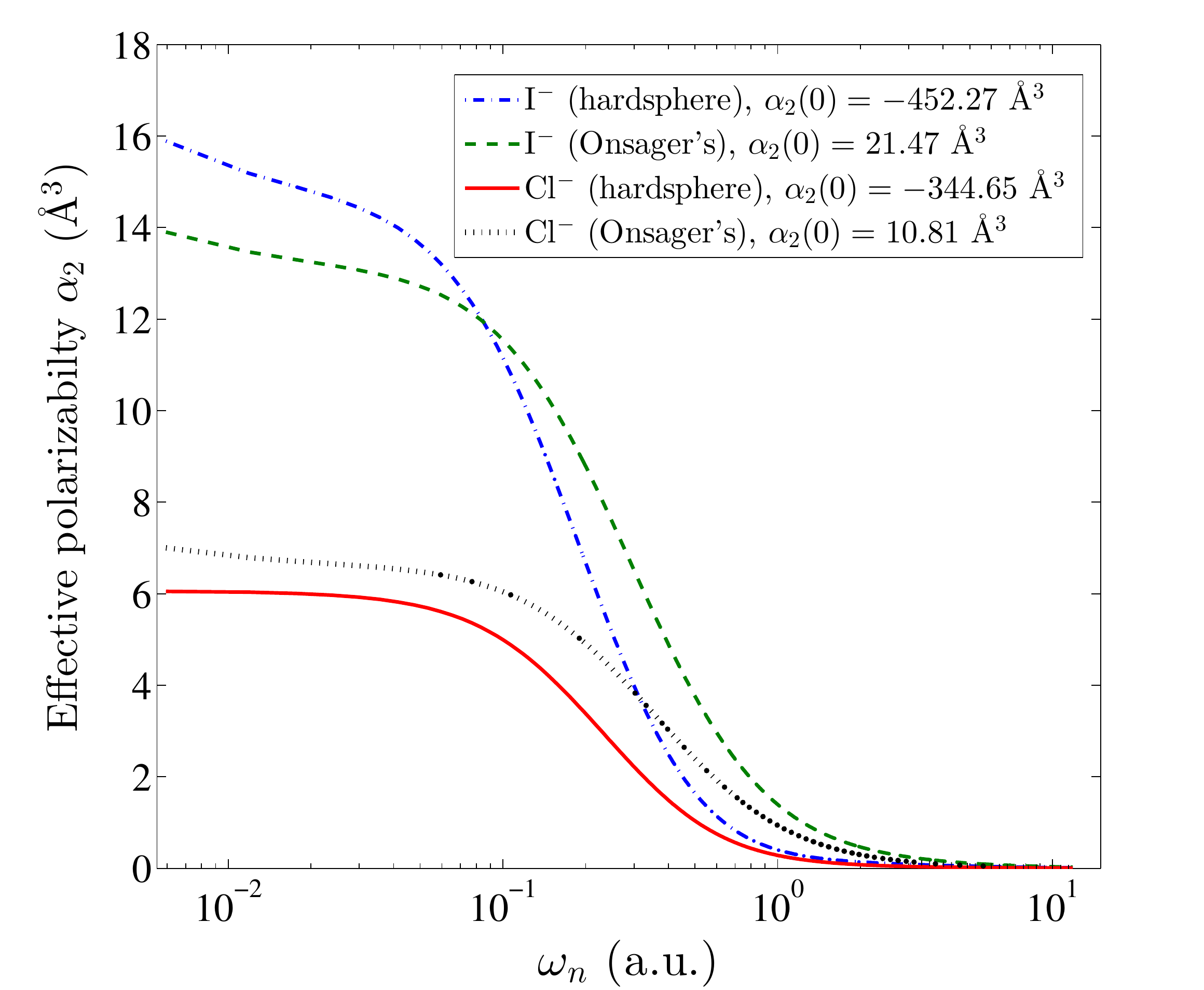}
\caption{(Color online) The effective polarizabilities of Cl$^{-}$ and I$^-$ for the Onsager and hardsphere models for discrete imaginary frequencies. The zero frequency values are shown in the figure legends.}
\label{figu1}
\end{figure}

Using these dielectric functions, the Lifshitz {force} at temperature $T$ can be
calculated following the formalism in the literature~\cite{Sernelius,Maha}. 
There is a small correction to the Lifshitz {force} due to salt ions\,\cite{Maha}. However, the effect of screening on the {fluctuations} due to salt ions on the  zero frequency term in the Lifshitz {force} is very small when we consider salt concentrations around 1-30 mM and water sheet thicknesses around 1-30 nm. 
 Ions in the water layer may also change the frequency-dependent part of the
dielectric response of the water layer, and thus may affect directly non-zero frequency contributions to the Lifshitz {force}. However, at the very low concentrations studied, the change in dielectric function is very small. 
Frequency intervals where the intervening
medium has a dielectric permittivity in between the permittivity of
the two surfaces holding the medium give a repulsive contribution;
other intervals give an attractive contribution. The calculation of
the {force} needs accurate dielectric functions over a wide range of
frequency for precise predictions.

%\paragraph{Ionic dispersion potentials.}
In order to explore the ion specific double layer interaction in an
uncharged three-layer system, we require the
%non-electrostatic
dispersion potential between ions and the different interfaces. We use
the full theory from Sambale et al.~\cite{Sambale09} to deduce the
simpler non-retarded expressions:

%3
\begin{equation}
U_i (z) = {\frac {B_{12}} {z^3}}+ {\frac {B_{32}} {(L-z)^3}},
\label{Eq3}
\end{equation}

%4
\begin{equation}
B_{j2} = {{\frac {k_B T } {2}} } \sum\limits_{n = 0}^{\infty}{'} 
{\frac{\alpha_2(i \omega_n)} {\epsilon_2(i \omega_n)}} \bigg ({ \frac{\epsilon_2(i \omega_n) -\epsilon_j(i \omega_n) } { \epsilon_j(i \omega_n) +\epsilon_2(i \omega_n)}\bigg )},
\label{Eq4}
\end{equation}
for a polarizable particle in the water phase at distances $z$ and
$L-z$ from the first and the second interfaces respectively. 
In fig.~\ref{figu1}, we show illustrative curves for the two models of effective
polarizability, viz., the Onsager and the hardsphere models for Cl$^{-}$ and I$^-$ ions. These are used to calculate ionic dispersion potentials acting between ions and interfaces. The evaluated $B_{12}$ values used to calculate ionic dispersion potentials are then given in table 1.
Here, $k_B$ is the Boltzmann constant, $T=273.16$ K is the
temperature, and $\alpha_2(i\omega_n )$ is the effective molecular
polarizability in water~\cite{Ninhb}. The prime indicates that the
$n=0$ term should be divided by 2. $\epsilon_1$, $\epsilon_2$ and
$\epsilon_3$ are the frequency-dependent dielectric functions of the
first material, water and the second material respectively at the
imaginary Matsubara frequencies $(i\omega_n)$ given by $\omega_n=2 \pi
k_B T n/\hbar$~\cite{Ninhb,Sernelius}.

 Two different models are
used for the effective polarizability $\alpha_2(\omega)$ of the salt
ions in water: (i)~Onsager's real-cavity model for local-field
corrections assumes that the particle is inside a small spherical
vacuum bubble embedded in the water medium~\cite{Onsager}. One finds
that~\cite{Sambale07,Sambale09}
%
%11
\begin{equation}
\alpha_{2}=\alpha\biggl(\frac{3\epsilon_2}{2\epsilon_2+1}\biggr)^2, \alpha(i\omega_n)=\alpha(0)\sum_j\frac{f_j}{1+(\omega_n/\omega_j)^2}
\label{Eq11}
\end{equation}
where $\alpha(i\omega_n)$  is the dynamic free-space polarizability~\cite{ParsonsNinham2010dynpol}. The adjusted parameters ${f_j}$ and ${\omega_j}$ are fitted to agree with the ab initio polarizability in vacuum ${\alpha(0)}$  by Parsons and Ninham~\cite{ParsonsNinham2010dynpol}. (ii)~The hardsphere model posits that
the particle is a homogeneous dielectric sphere of {radius} $a$. The {hard sphere ion radii are $a=$2.05, 2.16, 2.33, 0.67, and {1.06~\AA}} for Cl$^-$, Br$^-$, I$^-$, Na$^+$, and K$^+$~\cite{ParsonsNinham2009}. Its effective permittivity $\epsilon$ can be deduced from the free-space polarizability via~\cite{Jackson}
%
%13
\begin{equation}
\alpha=4 \pi \epsilon_0 a^3\,\frac{\epsilon-1}{\epsilon+2}.
\label{Eq13}
\end{equation}
The excess polarizability of the homogeneous sphere in water is then ~\cite{Sambale10}
%
%14
\begin{equation}
\alpha_2=4 \pi \epsilon_0 \epsilon_2a^3\,\frac{\epsilon-\epsilon_2}
 {\epsilon+2\epsilon_2}.
\label{Eq14}
\end{equation}

%\paragraph{Double-layer force.}
To study the specific ion effects, a calculation of the ion profiles
between the two interacting surfaces is required. The distribution of
ions between the surfaces is determined by a modified
Poisson-Boltzmann equation that includes both electrostatic and
nonelectrostatic potentials acting on the ions~\cite{Lima}
%17
\begin{equation}
{\frac {d^2\phi} {dz^2}}=-{\frac {e \sum_i z_i c_{i,0} \exp[-(z_i e \phi(z)+U_i (z))]} {\epsilon_0 \epsilon_w (0)}},
\label{Eq17}
\end{equation}
%18
\begin{equation}
{\frac {d\phi} {dz}}_{z=0}={\frac {d\phi} {dz}}_{z=L}=0.
\label{Eq18}
\end{equation}

The ionic dispersion potentials {are complemented}
 with a hardsphere repulsion, which prevents all the ions
from coming closer than one hardsphere radius of the ion from each
surface. The above expression together with a particular salt
concentration, water sheet thickness, $B_{j2}$ values  and the {hardsphere ion radii are} used to calculate the double
layer contribution~\cite{Lima},
%19
\begin{equation}
\begin{array}{l}
A(L)={\frac {e} {2}} \int_0^L \phi \sum_i c_i z_i dz+\int_0^L \sum_i c_i U_i dz+ \\
\quad  {\frac {1} {\beta }} \int_0^L \sum_i  [c_i \ln(c_i/c_{i,0})-(c_i -c_{i,0} )]dz,
\end{array}
\label{Eq19}
\end{equation}
where $A(L)$ is the free energy of two planar interfaces a distance $L$ apart, c$_i$ is the concentration of ion $i$ at a specific point and c$_{i,0}$ the ion concentration in the bulk. The {force} is then calculated as $p(L)=-dA(L)/dL$.

\begin{figure}
\includegraphics[width=8cm]{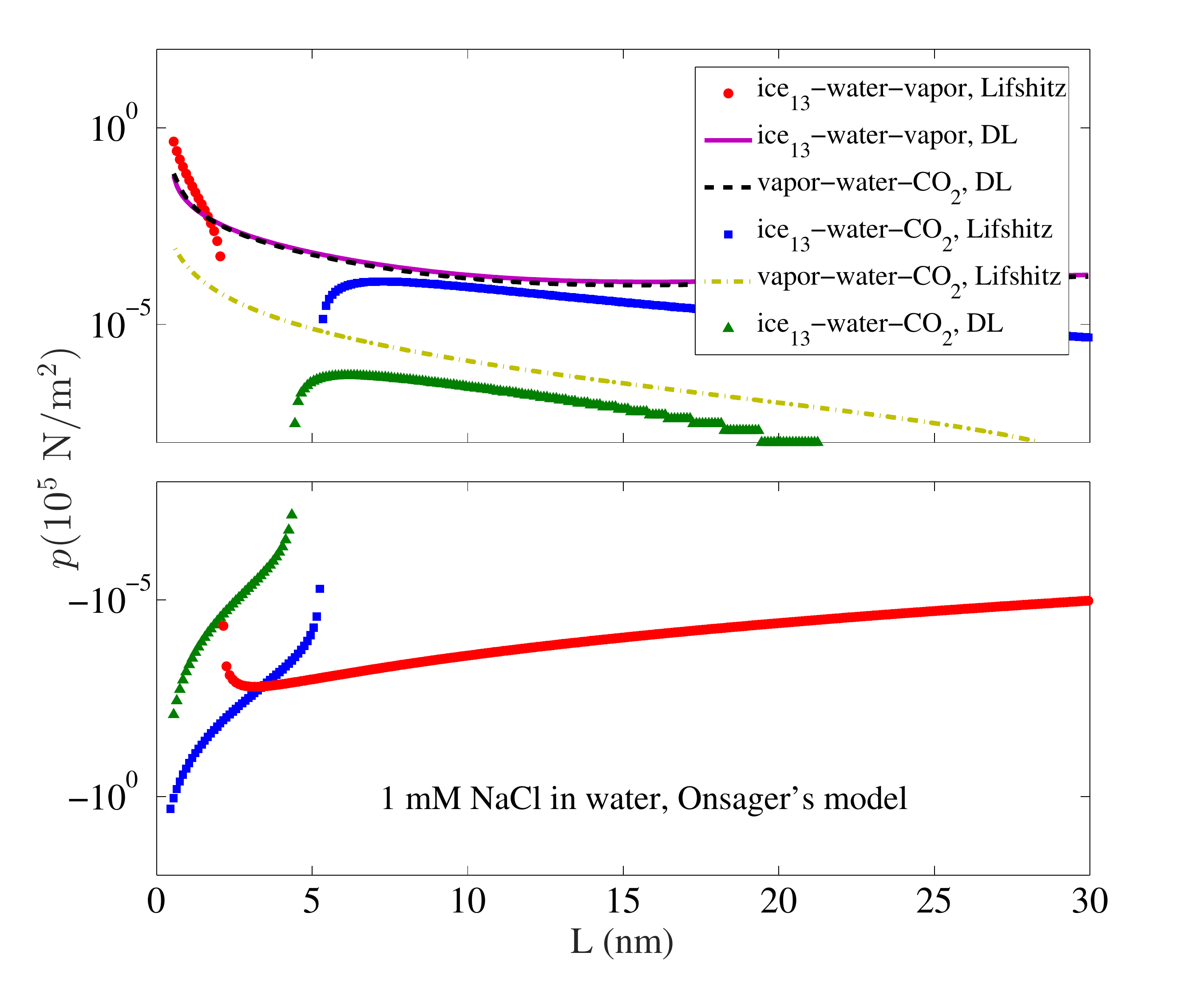}
\caption{(Color online) {The double layer force between two interfaces separated by  1 mM NaCl,  ice-water-vapor, ice-water-CO$_2$ and vapor-water- CO$_2$. The contribution to the total force arising from the Lifshitz energy is also shown.} (We use the ice$_{13}$ model for the dielectric function of ice and Onsager's model for the effective polarizability of salt ions in water). }
\label{figu2}
\end{figure}

\begin{table}
\caption{The $B_{12}$ values (the non-retarded van der Waals energy times $z^3$ in units of 10$^{-50} $J m$^3$) for ion-surface interaction for the hardsphere and Onsager's models near different interfaces. For the ice-water interface, we give two sets of $B_{12}$ values using the ice$_{13}$ and ice$_{12}$ (in parenthesis) models of dielectric function for ice~\cite{Elbaum}. }
\centering
\small\addtolength{\tabcolsep}{-3pt}
\label{values}
\begin{tabular}{cccc}
\hline
\hline
\multicolumn{1}{c}{Interface} &\multicolumn{1}{c}{Ionic species} &\multicolumn{1}{c}{$B_{12}$ (Onsager)} &\multicolumn{1}{c}{$B_{12}$ (hardsphere)}\\
\hline
                & Cl$^-$ &0.046 (-0.026)& 0.19 (0.18)\\
                &Br$^-$ &0.088 (0.0012)&   0.28 (0.27)\\
 ice-water&I$^-$&0.19 (0.089)& 0.48 (0.47)\\
                & Na$^+$ &-0.049 (-0.069)& -0.046 (-0.065) \\
                &K$^+$&-0.12 (-0.17)&  -0.081 (-0.12)\\
\hline
                   &Cl$^-$&13.50&  7.99 \\
                   &Br$^-$&17.14&  10.85\\
vapor-water&I$^-$&23.49&  16.47\\
                   &Na$^+$&0.82& 0.69 \\
                   &K$^+$&3.83& 3.45\\
\hline
                       &Cl$^-$&-5.70&  -3.56 \\
                       &Br$^-$&-7.17& -4.68 \\
CO$_2$-water&I$^-$&-9.67& -6.72\\
                       &Na$^+$&-0.45& -0.40 \\
                       &K$^+$&-1.87& -1.7\\
\hline
\hline
\end{tabular}
\end{table}

%\paragraph{Results.}
In fig.~\ref{figu2}, we show the ion-specific {total force for two interfaces separated by}
1 mM NaCl salt concentration in water and the Lifshitz {contribution to the total force} as a
function of distance across the salt water. {Here we consider
ice--water--vapor, ice--water--CO$_2$, and
vapor--water--CO$_2$ systems}. For systems without
salt, the double layer force is absent, and the force is given by the
retarded Lifshitz force. The double-layer contribution correlates with
the magnitude and sign of B$_{12}$ values at the different interfaces.
Both systems involving vapor--water interfaces are dominated by the
rather large repulsive B$_{12}$ values at this interface. The {force}
changes sign at 2.1 nm for the ice$_{13}$--water--vapor system for
Lifshitz interaction while no reversal of sign occurs for the double
layer interaction. The sign does not reverse in the
vapor--water--CO$_2$ system for both Lifshitz and the double layer
interactions.  {The ice--salt-water--CO$_2$ system} is
quite unique with a  sign reversal {in the force}  for both the linear
retarded Lifshitz term (at 5.3 nm) and the non-linear van der Waals
{force} from ion-surface interactions (at 4.4 nm). For 1 mM NaCl
concentration, we observe that the CO$_2$ bubbles in water tend to get
attracted to the ice surface when they come closer to the surface,
while they are always repelled from the vapor surface. The two
additional terms that also arise when we solve the Poisson-Boltzmann
equation for the ion profiles under the influence of van der Waals
forces and electrostatic forces are the entropic term and the
electrostatic term. The electrostatic term is almost negligible in the
present case, while the entropic term contributes to attraction for all
separations but the behaviour is
globally dominated by the non-linear van der Waals term. This term comes from ion-surface van der Waals force acting on the ions in the ion
profile. Similar to the linear Lifshitz term it contributes to short
range attraction and long range repulsion. {This means that the
CO$_2$ bubbles} that come very close to the interface
bind to the ice substrate by the combined effect of attractive linear
and non-linear dispersion forces.

\begin{figure}
\includegraphics[width=8cm]{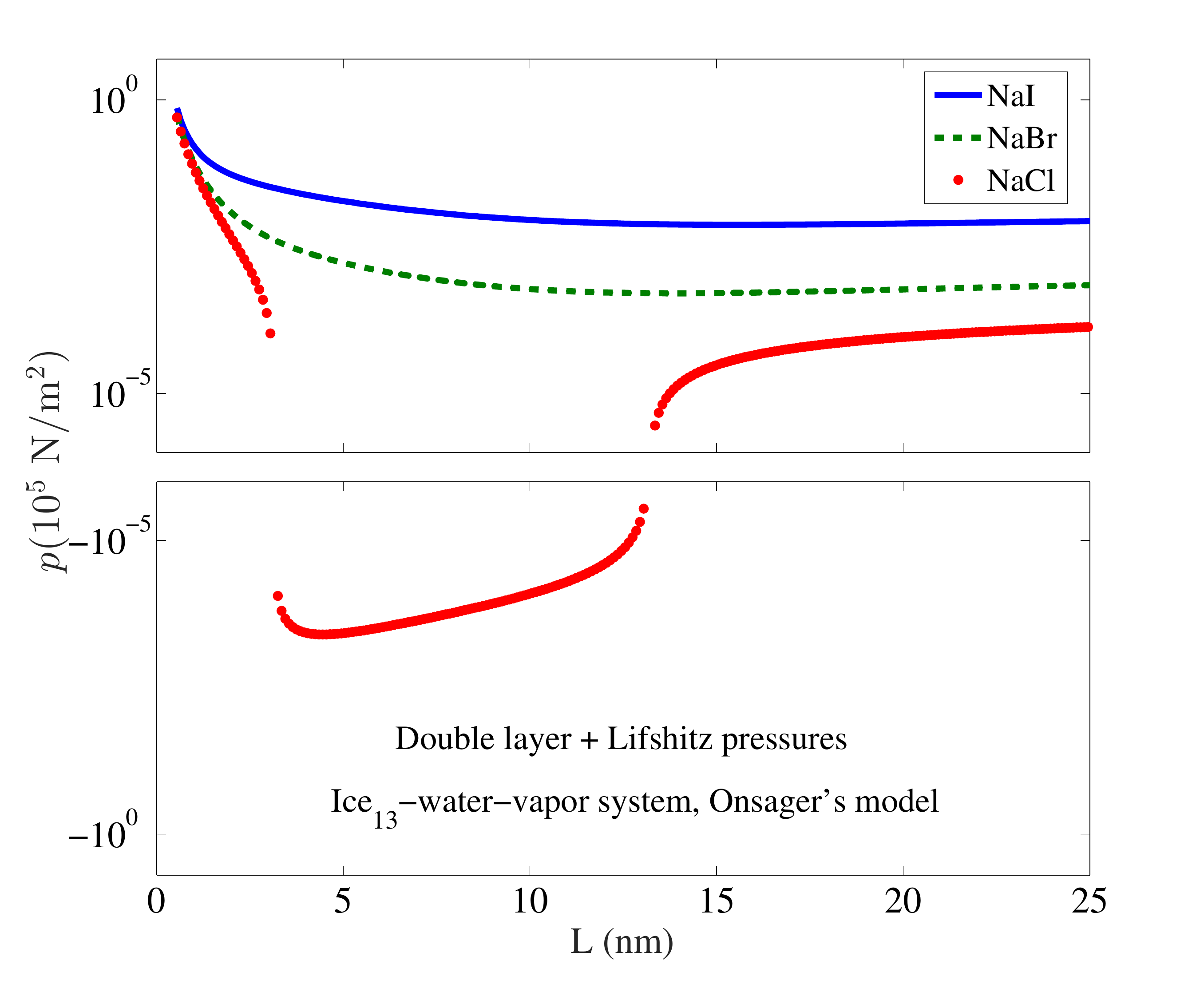}
\caption{(Color online) {The force between ice and vapor  interfaces separated by } 1 mM of different salt solutions as a function of water film {thickness}. We consider the ice$_{13}$ model for the dielectric function of ice  and the Onsager model for effective polarizability of the ions in water.}
\label{figu3}
\end{figure}

In fig.~\ref{figu3}, we present the {total force for the ice-water-vapor system} with a fixed
concentration (1 mM) of different salts in water using Onsager's
model of effective polarizability. The contribution from gravity
($-{\rho}gL$) is negligible for the distances considered here. The
reversal of the sign of {the force} in ice-water-vapor system facilitates
the formation of a thin film of water of finite thickness, otherwise
also known as incomplete surface melting~\cite{Elbaum}. We observe
from fig.~\ref{figu3} that reversal depends not only on the dielectric
functions of the three layers, but also on the ionic species. For
salts, like NaBr and NaI with higher values of positive van der Waals
$B$ parameters, the double layer {force} is too repulsive in the case
of Onsager's model so that there is no finite equilibrium water film
thickness. 1 mM NaBr and NaI give repulsion for all water film
thicknesses. In other words, with an unlimited reservoir of salt ions
in water surface melting of ice is complete.
If, however, there is a fixed amount of salt ions on the surface the
salt concentration will go down when the water sheet thickens. This
will stop the system from undergoing complete melting. An equilibrium thin film does appear for these ions with the hardsphere model.

\begin{figure}
\includegraphics[width=8cm]{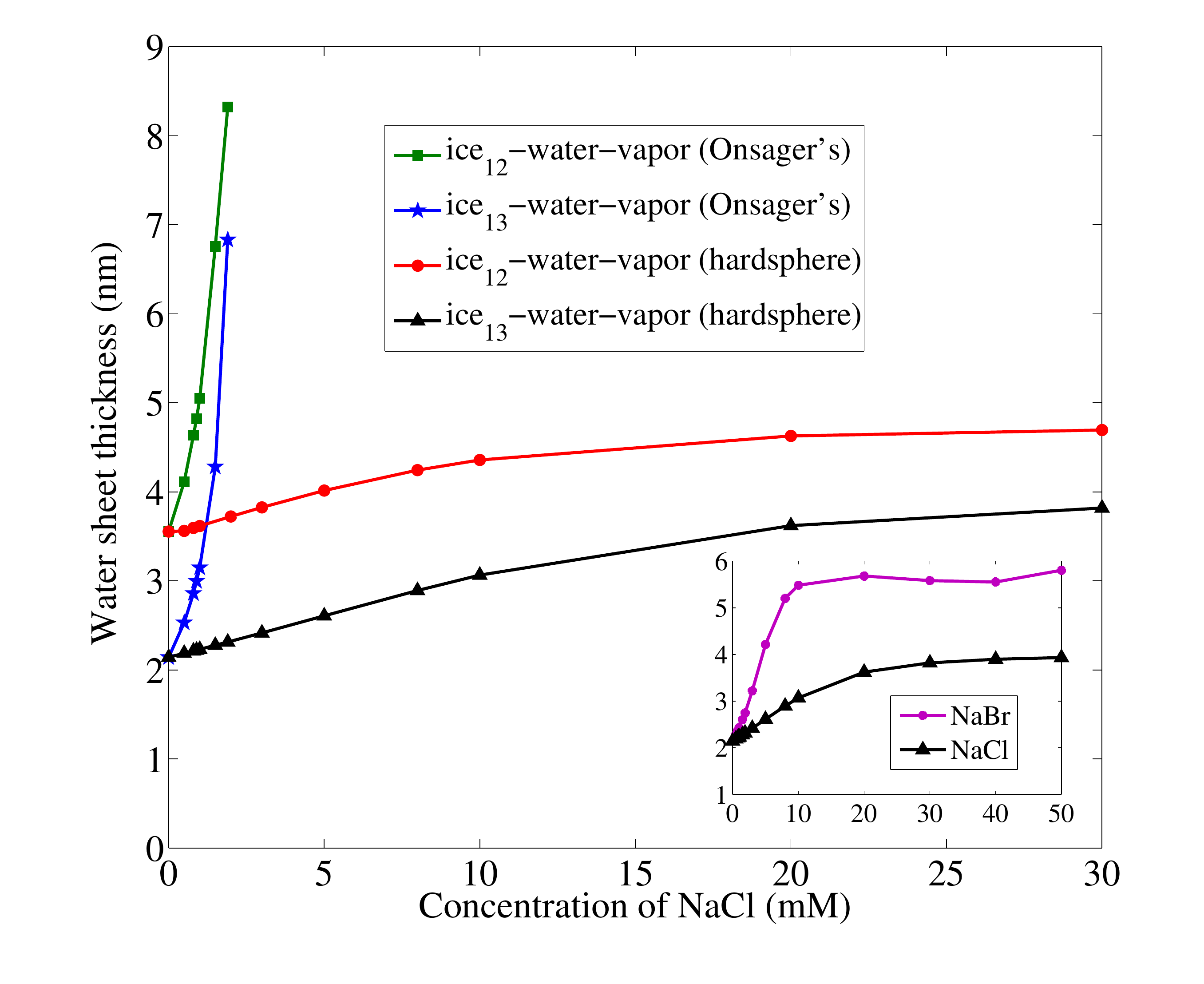}
\caption{(Color online)  Equilibrium water sheet thickness for ice-water-vapor system as a function of NaCl salt concentration using different models for the dielectric function of ice and the polarizability of the ions. In the inset figure, we compare the equilibrium water film thickness for different salts using the hardsphere model and the ice$_{13}$ model for the dielectric finction of ice.}
\label{figu4}
\end{figure}

In fig.~\ref{figu4}, we show that the equilibrium thickness of water
in the ice-water-vapor system increases with increasing concentration
of NaCl. The results change slightly with the application of different
models of the dielectric function of ice. However, the change due to
application of the different models of effective polarizability of
salt ions is significant. As stated earlier, Onsager's model for
effective polarizability gives faster growing water films than as
given by the hardsphere model. With the hardsphere model, the water
film first grows with concentration and then, maintains a constant
finite equilibrium thickness after a certain point. The equilibrium
water film grows faster with increasing NaBr concentration than with
increasing NaCl concentration (see inset of fig.~\ref{figu4}).

Our model neglects some microscopic aspects that are discussed in
the literature on ice premelting\,\cite{Limmer} (and references therein), such as the molecular organization of liquid water at
the interfaces (liquid-solid and liquid-vapor), that extends for 1-2 nm
from the surface, the consequent change of dielectric constant there,
and the effect of ions on them. For example, it is not always the case that
ions have no specific adsorption behavior at an interface mediated by
such molecular organization, and that this may be different for the
positive and negative ions making the electrostatic potential
different and the double-layer contribution to the thickness also
different.

In this work, we have explored a simple model system where
cold {bubbles} interact with ice--water
or vapor--water interfaces. This enables us to study the trends when van der Waals and Lifshitz forces are included. The vital retardation effects have not been accounted for in molecular simulations on premelting (see for instance the work of Limmer and Chandler\,\cite{Limmer}).
We have also investigated how the presence
of salt ions influences the equilibrium
thickness
of the Casimir-induced water layer on ice surface. We predict an
equilibrium water film thickness for ice-water-vapor system that
increases with the increase in concentration of salt in the water
phase. The results are directly applicable to the melting of ice under
equilibrium conditions discussed by Elbaum and Schick~\cite{Elbaum}.
We have found that the predictions strongly depend on the model
employed for the effective polarizability of the ion in water. We have used the Onsager and hardsphere models. For this reason, a
more realistic hybrid model as suggested in Ref.~\cite{Sambale10}
should be developed in the future, and this model should be
cross-compared with microscopic simulations based on
density-functional theory.{ One should note that we are neglecting the chemical equilibrium of CO$_2$ in water. 
This gas is not inert for water as it forms different hydrated species and ions. And its solubility depends on the ionic strength.
This may very well influence the molecules and ion concentration at the interface and in solution. We plan to return to these interesting aspects in a later publication which will also include the effect of varying the pH in solution.}

%\begin{acknowledgments}
PT acknowledges support from the European Commission; this publication
reflects the views only of the authors, and the Commission cannot be
held responsible for any use which maybe made of the information
contained therein. SYB gratefully acknowledges support by the German
Research Foundation (grant BU 1803/3-1) and the Freiburg Institute for
Advanced Studies.
MB, OIM, and CP acknowledge support from the Research Council of Norway (Project: 221469). ERAL acknowledges support from the Brazilian agency FAPERJ.  We acknowledge access to HPC resources at NSC through SNIC and at USIT through NOTUR. We thank Prof. B. W. Ninham for his comments on our manuscript.

%\end{acknowledgments}

\end{document}